\documentclass[a4paper, 11pt]{article}
\usepackage[english]{babel}
\makeatletter

\usepackage{geometry,graphicx,color}
\usepackage{amsfonts, amsmath, amssymb, slashed,dsfont}
\usepackage{tensor}
\usepackage{hyperref}
\usepackage{epsfig}
\usepackage{pifont}
  \usepackage{soul}
\usepackage{enumitem}
\usepackage{csquotes}
\usepackage{bm}
\usepackage{mathrsfs} 
\usepackage[hyperref,thmmarks,amsmath]{ntheorem}
\usepackage{amsmath,amssymb,slashed}
\usepackage{comment}
\usepackage{tikz}
\usetikzlibrary{positioning}
\usetikzlibrary{decorations.markings}
\usetikzlibrary{arrows.meta}
\usepackage{filecontents}
\usepackage{comment}
\numberwithin{equation}{section}
\newcommand\bbone{\mathbb{I}}

\DeclareMathOperator{\tr}{Tr} 

\newcommand\sstar{\text{\ding{86}}}

\newcommand{\institute}[1]{\newcommand{\@institute}{#1}}
\renewcommand{\maketitle}{
\vspace*{0.5\baselineskip}
{
\center\LARGE\noindent\@title\par
}%
\vspace{1.5\baselineskip}
{
\center\normalsize\noindent\ignorespaces\@author\par
}%
\vspace{0.5\baselineskip}
{
\center\normalsize\ignorespaces\@institute\par
}%
\vspace{2\baselineskip}
}%

\usepackage{siunitx}
\usepackage{xcolor}
\usepackage{booktabs,colortbl, array}
\usepackage{pgfplotstable}
\usepackage{pbox}
\definecolor{rulecolor}{RGB}{0,71,171}
\definecolor{tableheadcolor}{gray}{0.92}
\newcommand{\topline}{ %
        \arrayrulecolor{rulecolor}\specialrule{0.1em}{\abovetopsep}{0pt}%
        \arrayrulecolor{tableheadcolor}\specialrule{\belowrulesep}{0pt}{0pt}%
        \arrayrulecolor{rulecolor}}
\newcommand{\midtopline}{ %
        \arrayrulecolor{tableheadcolor}\specialrule{\aboverulesep}{0pt}{0pt}%
        \arrayrulecolor{rulecolor}\specialrule{\lightrulewidth}{0pt}{0pt}%
        \arrayrulecolor{white}\specialrule{\belowrulesep}{0pt}{0pt}%
        \arrayrulecolor{rulecolor}}
\newcommand{\bottomline}{ %
        \arrayrulecolor{white}\specialrule{\aboverulesep}{0pt}{0pt}%
        \arrayrulecolor{rulecolor} %
        \specialrule{\heavyrulewidth}{0pt}{\belowbottomsep}}%

\pgfplotstableset{normal/.style ={%
        header=true,
        string type,
        column type=l,
        every odd row/.style={
            before row=
        },
        every head row/.style={
            before row={\topline\rowcolor{tableheadcolor}},
            after row={\midtopline}
        },
        every last row/.style={
            after row=\bottomline
        },
        col sep=&,
        row sep=\\
    }
}

\begin{document}

\title{Star-products for Lie-algebraic noncommutative Minkowski space-times }
\author{Valentine Maris$^{a,b}$, Filip Po\v{z}ar$^c$, Jean-Christophe Wallet$^a$}
\institute{%
\textit{$^a$IJCLab, Universit\'e Paris-Saclay, CNRS/IN2P3, 91405 Orsay, France}\\
\textit{$^b$Laboratoire de Physique, ENS de Lyon, Universit\'e de Lyon,\\
CNRS UMR5672, 69007 Lyon, France}\\
\textit{$^c$Rudjer Bo\v{s}kovi\'c Institute, Bijen\v{c}ka c.54, HR-10002 Zagreb, Croatia}
\bigskip\\

e-mail:  
\href{mailto:valentine.maris@ijclab.in2p3.fr}{\texttt{valentine.maris@ens-lyon.fr}}\\
\href{mailto:filip.pozar@irb.hr}{\texttt{filip.pozar@irb.hr}}\\
\href{mailto:jean-christophe.wallet@universite-paris-saclay.fr}{\texttt{jean-christophe.wallet@universite-paris-saclay.fr}}
}%

\maketitle

\begin{abstract} 

Poisson structures of the Poincar\'e group can be linked to deformations of the Minkowski space-time, classified some time ago by Zakrewski. Based on this classification, various quantum Minkowski space-times with coordinates Lie algebras and specific Poincare Hopf algebras have been exhibited by Mercati and called T-Minkowski space-times". Here we construct the star products and involutions characterizing the $\star$-algebras for a broad family of Lie algebras which includes 11 out of 17 Lie algebras of T-Minkowski spaces. We show that the usual Lebesgue integral defines either a trace or a KMS weight (”twisted trace”) depending on whether the Lie group of the coordinates' Lie algebra is unimodular or not. Finally, we give the Poincaré Hopf algebras when they are compatible with our $*$-product. General derivation of such symmetry Hopf algebras are briefly discussed.

\end{abstract}
\vfill\eject
\section{Introduction}\label{section1}

Quantum deformations of the Minkowski space-time have received considerable attention for more than three decades as they are consensually believed to have a promising physical interest, being of possible relevance in a description of an effective regime of Quantum Gravity \cite{zerevue}, \cite{whitepaper}. Among these quantum space-times, the $\kappa$-Minkowski space-time became from year to year the most popular one in the physical literature \cite{luk-rev} due to some of its salient properties, for instance as providing a realization of the Double Special Relativity \cite{kow-glik} or its possible relationship to Relative Locality \cite{rel-loc1}, \cite{rel-loc2}. Its coordinates algebra is a Lie algebra given by $[x_0,x_i]=\frac{i}{\kappa}x_i$, $[x_i,x_j]=0$, for $i,j=1,\cdots, (d-1)$, where $\kappa$ is the deformation parameter with mass dimension $-1$.
Recall that the associative algebra describing the $\kappa$-Minkowski space-time, denoted by $\mathcal{M}_\kappa$, is acted on by a deformation of the Poincar\'e algebra, denoted by $\mathcal{P}_\kappa$, and is (Hopf) dual to the translations Hopf subalgebra of $\mathcal{P}_\kappa$ so that the entities $\mathcal{M}_\kappa$ and $\mathcal{P}_\kappa$ are rigidly linked together, roughly as a pair involving a quantum space-time and its "quantum isometries".\\

 Other Lie-algebraic deformations of the Minkowski space-time have also been identified for a long time but have not been intensively considered so far. These quantum Minkowski space-times are linked to different quantum deformations of the Poincar\'e/relativistic symmetry, in a similar way as $\mathcal{M}_\kappa$ and $\mathcal{P}_\kappa$ are linked as recalled above. It turns out that these deformations are related to the Poisson structures of the Poincar\'e group given by classical $r$-matrices, which have been classified a long ago in \cite{zakrzew}, see also \cite{Z2}. This result has been used in \cite{lukier1} to construct many Lie-algebraic deformations of the Minkowski space-time, whose coordinates Lie algebra has the general form 
\begin{equation}[x^\mu,x^\nu]=i\zeta^\mu(\eta^{\mu\beta} x^\alpha-\eta^{\nu\beta} x^\alpha)-i\zeta^\nu(\eta^{\mu\beta} x^\alpha-\eta^{\nu\beta} x^\alpha),
\end{equation}
where $\zeta^\mu$ is a vector with dimension of a lenght and $\alpha$ and $\beta$ fixed. Among this family of Lie algebras is the so called $\rho$-Minkowski space-time\footnote{Its Lie algebra of coordinates is given by $[x_0, x_1] = i \rho x_2,\ [x_0, x_2] = - i \rho x_1,\ [x_1, x_2] = 0,$ where the deformation parameter $\rho$ has the dimension of a lenght}, recovered for $\zeta^\mu=\delta^\mu_0 $, $\alpha=1$, $\beta=2$, which has been considered in the literature some times ago from various viewpoints, see \cite{localiz1}-\cite{gauge-rho}. \\

An attempt to extend the above analysis has been recently carried out in \cite{mercati}, based on a few reasonable assumptions which amounts to require the existence of triangular $R$-matrices in one-to-one correspondence with the $r$-matrices of the Poincar\'e algebra classified in \cite{zakrzew}. This gives rise \cite{mercati} to 17 different classes of centrally-extended Lie algebras of coordinates characterizing these new quantum Minkowski space-times.\\

The purpose of the present paper is to provide a systematic construction of star products and involutions for these 17 models thus defining the corresponding noncommutative algebras describing these quantum Minkowski space-times. We will restrict ourselves for each of the above 17 classes to the simple case of non centrally-extended Lie algebras of coordinates which nevertheless leads to the characterization of 10 new quantum Minkowski space-times with "noncommutativity of Lie-algebra type". \\

In this paper, the star-products defining the various deformations of the Minkowski space-time will be obtained from a generalisation of a construction introduced a long time ago by von Neumann \cite{vonNeumann} in order to formalize the work of Weyl on the phase space quantization \cite{Weyl} which has led to the definition of the popular Moyal product. This construction proves very convenient whenever the space-time non-commutativity is of "Lie algebra type" and has already been exploited in the context of $\kappa$-Minkowski and $\rho$-Minkowski, \cite{rho-weyl}, \cite{kappa-weyl}-\cite{PW2019}. Schematically, it uses the defining properties of convolution algebras for the group $\mathcal{G}$ related to the Lie algebra of coordinates $\mathfrak{g}$ combined with Weyl quantization map, hereafter denoted by $Q$.
The benefit is that it provides altogether both a star-product expressed as an integral convenient for practical computations, an involution and a natural trace characterizing a well defined associative $*$-algebra modeling a deformed Minkowski space-time, generically denoted hereafter by $\mathcal{M}$.\\

\noindent The paper is organized as follows: \\
In Section \ref{section2}, the construction for the star-products is presented. A general formula is given and discussed. We also provide the formulas for the corresponding involutions. \\
In Section \ref{section3}, we apply the general results to each of the ten yet unexplored quantum Minkowski space-times, discussing each case in detail and in particular paying attention to the case of non-unimodular groups related to some of the coordinate algebras for which the notion of trace must be replaced by a KMS weight as it already happens in the popular $\kappa$-Minkowski case \cite{kappa-weyl}, \cite{kappa-weyl-bis}. \\
Section \ref{section4} deals with the characterization of the deformed relativistic symmetries acting on the quantum Minkowski space-times. These (roughly speaking) "quantum isometries" are described in each case by a deformed Hopf algebra, say $\mathcal{H}$,  such that $\mathcal{M}$ is a left-module algebra over $\mathcal{H}$.\\
In Section \ref{section5}, we summarize the results and conclude.

\section{Quantum Minkowski space-times and related algebras}\label{section2}
\subsection{Weyl quantization map: from convolution algebras to star-products}\label{section21}

The whole formalism has been used and commented in detail in e.g \cite{gauge-rev}. To make the paper self-contained, this subsection summarizes the main steps of the construction and collects the corresponding mathematical tools. \\
Notice that in this section we will set all the dimensionful constants equal to $1$ in order to simplify the notations. These will be reinstalled in Section \ref{section3}. \\
Notice also that the present construction can describe quantum space-times with "noncommutative time" or "commutative time", depending on the initial choice for the labeling of the (momentum) coordinates. Passing from one case to the other is straightforward. In this section, we will present the results in their most general form with the coordinate $X^M$ being the main noncommutative coordinate which does not commute with other coordinates.\\

The first step amounts to determine the Lie group associated to each Lie algebra of coordinates. It turns out that each (locally compact) group considered in this paper is found to have a semi-direct product structure. Namely, one has
\begin{equation}
    \mathcal{G}
    = H\ltimes_{{\phi}} \mathbb{R}^{n}
    \label{genesemidirect}
\end{equation}
$n\ge1$, where $H\subset GL(n,\mathbb{R})$ is an abelian Lie group which depends on the structure of the Lie algebra $\mathfrak{g}$ as it will be illustrated in the next subsection and $\phi: H \to \mathrm{Aut}(\mathbb{R}^{n})$, a continuous group morphism, defines the action of $H$ on $\mathbb{R}^n$ viewed as additive groups. \\

In the ensuing analysis, it will be convenient to use a faithful representation of $\mathcal{G}$, $\gamma:\mathcal{G}\to \mathbb{M}_{n+1}(\mathbb{C})$
\begin{equation}
    \gamma:(a,x)
    \longmapsto \begin{pmatrix} a & x \\ 0 & 1   \end{pmatrix}\label{decadix-gamma}
\end{equation}
for any $a\in H$, $x\in\mathbb{R}^n$, from which one infers that the action of any element of $\mathcal{G}$ on $\mathbb{R}^n$ is represented by $\gamma((a,x))y=ay+x$ for any $y\in\mathbb{R}^n$ while $\phi$ in \eqref{genesemidirect} is given by $\phi_a(x) = a x$, for any $a\in \mathbb{M}_{n}(\mathbb{C})$, $x\in\mathbb{R}^n$, i.e. the standard action of the matrix $a$ on a vector $x$ of $\mathbb{R}^n$. Here, the pair $(a,x)$ denotes any element of $\mathcal{G}$. \\

The group law for $\mathcal{G}$ \eqref{genesemidirect} is given by
\begin{align}
    (a_1, x_1) (a_2, x_2)
    &= (a_1 a_2, x_1 + a_1 x_2), &
    \label{w1}\\
    (a,x)^{-1}
    &= (a^{-1}, -a^{-1} x), &
    \bbone_{\mathcal{G}}
    = (\bbone_H,0)
    \label{w2},
\end{align}
and will be explicitly used to obtain the convolution product.\\

The second step uses the main features of the convolution algebra for $\mathcal{G}$. \\
We denote by $d\mu_\mathcal{G}$ (resp. $d\nu_\mathcal{G}$) the left-invariant (resp. right-invariant) Haar measure on $\mathcal{G}$, with 
\begin{equation}
   d\nu_\mathcal{G}(s)=\Delta_\mathcal{G}(s^{-1})d\mu_\mathcal{G}(s)\label{modularfunction}
\end{equation}
for any $s\in\mathcal{G}$ where the group homomorphism $\Delta_\mathcal{G}:\mathcal{G}\to\mathbb{R}^+$ denotes the modular function. Recall that the convolution algebra for $\mathcal{G}$, $L^1(\mathcal{G})$, is a $*$-algebra equipped with the convolution product built with respect to the left Haar measure \footnote{An equivalent formulation of the convolution product can be build with respect to the right measure. Changing $t \to t^{-1}$ gives $d\mu_\mathcal{G}(t) \Delta(t^{-1}) F(s t^{-1}) G(t)$ and $d\mu_\mathcal{G}(t) \Delta(t^{-1}) = d\nu_\mathcal{G}(t)$ so we retrieve back \cite{kappa-weyl-bis}.} and involution defined as in \cite{folland2016course}
\begin{equation}
    (F\circ G)(s)
    = \int_{\mathcal{G}}\ d\mu_\mathcal{G}(t) F(s t) G(t^{-1})
    \label{convol-gene},
\end{equation}
\begin{equation}
    F^\sstar(u)
    = {\overline{F}}(u^{-1}) \Delta_\mathcal{G}(u^{-1})
    \label{involution}
\end{equation}
for any $F,G\in L^1(\mathcal{G})$, $s,t,u\in\mathcal{G}$, where ${\overline{F}}$ is the complex conjugate of $F$ and any group elements in \eqref{convol-gene}, \eqref{involution} are of the general form $(a,x)$ (see \eqref{decadix-gamma}). \\
For semi-direct products as given by \eqref{genesemidirect}, the Haar measure and modular function take the form 
\begin{eqnarray}
     d\mu_\mathcal{G}((a,x))
    &=& d\mu_{\mathbb{R}^n}(x)\ d\mu_H(a)\ |\det(a)|^{-1}
    \label{measure},\\
    \Delta_\mathcal{G}((a,x))
    &=& \Delta_{\mathbb{R}^n}(x)\ \Delta_H(a)\ |\det(a)|^{-1}
    \label{modul-function},
\end{eqnarray}
in obvious notation. Owing to the fact that $d\mu_{\mathbb{R}^n}(x)$ is the usual Lebesgue measure on $\mathbb{R}^n$ and $\Delta_{\mathbb{R}^n}(x)=1$ as the additive group $\mathbb{R}^n$ is unimodular, \eqref{measure}, \eqref{modul-function} take the form
\begin{eqnarray}
  d\mu_\mathcal{G}((a,x))
    &=& d^nx\ d\mu_H(a)\ |\det(a)|^{-1}
    \label{measure1},\\
    \Delta_\mathcal{G}((a,x))
    &=& \Delta_H(a)\ |\det(a)|^{-1}
    \label{modul-function1}, 
\end{eqnarray}
for all $a\in H$. As per our initial assumption \eqref{genesemidirect}, the group $H$ is abelian, and as such is unimodular\footnote{For the space 9 in \cite{mercati}, the group $H$ is given as $\mathbb{R}\ltimes \mathbb{R}$ which is not abelian and not unimodular. This is the main reason why we did not study the space 9. Generalization to the space 9 is possible, but the resulting convolutional $\star$ product would involve even more nested integrals.}, i.e.,
\begin{equation}
    \Delta_H(a)=1
\end{equation}
so that \eqref{modul-function1} simply reduces to 
\begin{equation}
   \Delta_\mathcal{G}((a,x))
    = |\det(a)|^{-1},  
\end{equation}
i.e., the unimodularity property of $\mathcal{G}$ is controlled by the value of $\det(a)$, $a\in H\subset GL(n,\mathbb{R})$. Accordingly, the left-invariant Haar measure \eqref{measure1} will simplify into
    \begin{equation}
        d\mu_\mathcal{G}((a,x))
    = d^nx\ dz\ |\det(a(z))|^{-1}\label{mesurefinale}
    \end{equation}
where the second Lebesgue measure $dz$ comes from the particular features of the subgroups $H$ as it will be apparent in a while from their explicit description. \\

Now as the third step, we assume that the functions of $L^1(\mathcal{G})$ are functions of the momentum space, that is, any $F\in L^1(\mathcal{G})$ can be written as $F=\mathcal{F}f(p)$ where $\mathcal{F}$ denotes the Fourier transform {\footnote{
    Our convention for the Fourier transform is $\mathcal{F}f(p) = \int \frac{d^dx}{(2\pi)^d}\ e^{- i p^M x^M - i \vec{p} \vec{x}} f(x)$ and for the inverse $f(x) = \int d^dp\ e^{i p^M x^M + i \vec{p} \vec{x}} \mathcal{F}f(p)$.
}} with $p=(p^M, \vec{p})$. So that any element of $\mathcal{G}$ can be conveniently parametrized as 
\begin{equation}
s(p^M,\vec{p})=\begin{pmatrix} a(p^M) & \vec{p} \\ 0 & 1   \end{pmatrix}\label{decadix1},
\end{equation}
stemming from \eqref{decadix-gamma} where  $a(p^M)$ is a matrix depending on one parameter identified with the energy if $p^M$ is time directed, momentum if $p^M$ space-like oriented and an appropriate linear combination of momenta and energy if $p^M$ is light like\footnote{Additionally, some spaces in time like and space like $p^M$ classes have boosted momentum as the generator of noncommutativity. In those cases, $p^M$ is interpreted as the boosted energy and momentum, respectively, for time like and space like $p^M$. This is further discussed in the Subsection \ref{section22}.}. The matrix $a(p^M)$ obviously satisfies
\begin{equation}
    a(p^M)a(q^M)=a(p^M+q^M) \label{matrix-exponent},
\end{equation}
\begin{equation}
a^{-1}(p^M)=a(-p^M)\label{flip-inverse},
\end{equation}
\begin{equation}
    a(0)=\bbone\label{adezero},
\end{equation}
for any $a\in H\subset GL(n,\mathbb{R})$. These relations generate some simplification in the general form of the star-products and their related involutions. They can be verified to hold for each of the deformations of the Minkowski space-time given in the next section.\\

By combining \eqref{w1}, \eqref{w2}, \eqref{mesurefinale} with \eqref{convol-gene}, one can rewrite the convolution product \eqref{convol-gene} and corresponding involution \eqref{involution} as
\begin{align}
(\mathcal{F}f\circ \mathcal{F}g)(p^M,\vec{p})
    = \int_{\mathcal{G}}\ d^n q\ dq^M\left|\det(a(q^M))\right|^{-1} 
\mathcal{F}f(p^M +q^M,\vec{p} +a(p^M)\vec{q})\mathcal{F}g(-q^M, -a^{-1}(q^M)\vec{q})
    \label{convol-1},
\end{align}
\begin{equation}
     \mathcal{F}f^\sstar(p^M,\vec{p})
    = {\overline{\mathcal{F}f}}(-p^M, -a^{-1}(p^M)\vec{p})\times \left|\det\left(a( p^M)\right)\right|,
\end{equation}
 where $a(p^M)\vec{p}$ denotes the action of any matrix $a(p^M)\in H$ on 
$\vec{p} \in\mathbb{R}^n$. As a technical remark, note that $L^1(\mathcal{G})$ has to be enlarged by a suitable completion. We will denote by $\mathbb{C}(\mathcal{G})$ the resulting group algebra whose explicit characterization will be not of our concern in the present paper.\\

As the last step of the construction, we introduce the Weyl quantization map $Q$ which associates to any function of a noncommutative $*$-algebra $\mathcal{M}$, equipped with star-product $\star$ and involution $^\dag$, a bounded operator on a Hilbert space $\mathcal{H}$ \cite{vonNeumann} \cite{Weyl}. Let $\mathcal{B}({\mathcal{H}})$ denote the corresponding operator algebra. As mentioned before, the algebra $\mathcal{M}$ models the noncommutative space with coordinates algebra 
$\mathfrak{g}$. The Weyl map is a $*$-algebra morphism defined as
\begin{equation}
    Q:\mathcal{M}\to\mathcal{B}({\mathcal{H}}),\ \ Q(f)
    = \pi(\mathcal{F}f)\label{Weyl-map},
\end{equation}
where the induced $\star$-representation of $\mathbb{C}(\mathcal{G})$ on $\mathcal{B}({\mathcal{H}})$, $\pi:\mathbb{C}(\mathcal{G})\to\mathcal{B}({\mathcal{H}})$, is defined by 
\begin{equation}
   \pi(F)
   = \int_\mathcal{G} d\mu_\mathcal{G}(x) F(x) \pi_U(x)
   \label{inducedrep} 
\end{equation}
for any $F(=\mathcal{F}f)\in\mathbb{C}(\mathcal{G})$ and where $\pi_U:\mathcal{G}\to\mathcal{B}({\mathcal{H}})$ is a unitary representation of $\mathcal{G}$. The quantization map $Q$ is a bounded and non-degenerate $*$-algebra morphism. Finally, the combination of \eqref{Weyl-map} with the fact that $Q$ and $\pi$ are both $*$-algebra morphisms yields
\begin{align}
    f\star g
    = \mathcal{F}^{-1} (\mathcal{F}f \circ \mathcal{F}g), && 
    f^\dag
    = \mathcal{F}^{-1} (\mathcal{F}(f)^\sstar)
    \label{star-prodetinvol},
\end{align}
for any $f,g\in\mathcal{M}$.\\

The above construction gives rise to an associative product and corresponding involution thus defining an associative $*$-algebra of functions $\mathcal{M}$ whose elements are interpreted as inverse Fourier transformed elements of $\mathbb{C}(\mathcal{G})$ introduced above. Thus, it really is natural to view $\mathcal{M}$ as an algebra of functions modeling the quantum (noncommutative) manifold 
with the Lie algebra of coordinates given by $\mathfrak{g}$.\\

A general expression describing the family of star-products given in \eqref{star-prodetinvol}
can be obtained from a standard computation by expressing the various Fourier transforms in \eqref{star-prodetinvol} combined with \eqref{convol-1}. In particular, integrating upon $d^nq$ generates a delta function appearing in the complete expression as 
$$...\int d^nz\ \delta\Bigl((a(q^M)^{-1})^T\vec{z}- a(p^M)^T\vec{x}\Bigr)\times...  $$ 
where the superscript $^T$ denotes the matrix transposition. Then, the integration over $d^nz$ (the integration variable in $\mathcal{F}f$) generates an overall factor $|\det((a(q^M)^{-1})^T)|^{-1}=|\det(a(q^M)|$ which balances the factor $|\det(a(q^M)|^{-1}$ involved in \eqref{convol-1} so that the resulting expression does not depend whether or not the group is unimodular as it does not depend on $\det(a(q^M))$. Introducing a diagonal extension of the matrix $a^T(p^M)$,
\begin{equation}
    A(p^M) = \mathbb{I}^M \oplus a^T(p^M) \oplus^{4 - n-1} \mathbb{I}
\end{equation}
one finally obtains
\begin{equation}
    f\star g = \frac{1}{(2\pi)} \int dp^M dy^M e^{-i p^M y^M} f(x + y^M)g(A(p^M)x)
\label{final star-product}
\end{equation}
for any $f,g\in\mathcal{M}$, where the argument of $f$ needs to be understood as vector addition 
\begin{equation}
    x + y^M \equiv x^\mu+ \delta^\mu\;_M y^M.
\end{equation}
A similar computation leads to the general expression for the involution. It is given by
\begin{equation}
    f^\dag(x_0,\vec{x})=\frac{1}{2\pi}\int dp^Mdy^M\ |\det(A(p^M))|^{2}e^{-ip^M y^M}\overline{f(A(p^M)x + y^M)}\label{gene-invol},
\end{equation}
which turns the associative algebra $\mathcal{M}$ into a $*$-algebra.\\

At this stage, one comment is in order. The expression \eqref{involution} for the involution equipping 
$L^1(\mathcal{G})$ insures that $\pi$ \eqref{inducedrep} defines a $*$-morphism. Indeed, one can also define the involution as $F^\sstar(u)
    = {\overline{F}}(u^{-1}) \Delta_\mathcal{G}(u^{\alpha})$, $\alpha\in\mathbb{Z}$ in a well defined manner, but the following relation is independent from the involution definition
    \begin{equation}
        \langle u, \pi(F)^\dag v\rangle=\langle \pi(F) u, v\rangle=\int_\mathcal{G} d\mu(s)\bar{F}(s^{-1})\Delta_\mathcal{G}(s^{-1})
        \langle u, \pi_U(s) v\rangle\label{exp1}
    \end{equation}
   for any $u,v\in\mathcal{H}$, where we used the linearity of the Hilbert product
$\langle .,.\rangle$ and the fact that $\pi_U$ is a unitary representation together with $d\mu(s^{-1})=\Delta_\mathcal{G}(s^{-1})d\mu(s)$ for any $s\in\mathcal{G}$. But on the other hand, from the new expression of the involution given above, one would infer
\begin{equation}
\langle u, \pi(F^*) v\rangle=\int_\mathcal{G} d\mu(s)\bar{F}(s^{-1})\Delta_\mathcal{G}(s^{\alpha})\langle u, \pi_U(s) v\rangle\label{exp2}
\end{equation}
so that $\pi(F)^\dag=\pi(F^*)$ is verified whenever $\alpha=-1$ \footnote{Notice that the value of $\alpha$ depends on the choice of left or right Haar measure. In \cite{kappa-weyl} \cite{kappa-weyl-bis} the right Haar measure is used, imposing $\alpha = +1$. Expression of the involution will depend of $\alpha$. In the case $\alpha = + 1$ the $|\det A(p^M)|^{2}$ factor is removed from \eqref{gene-invol}.}.

\subsection{Traces and commutation relations among coordinates}\label{section22}
Upon using the star product given in \eqref{star-prodetinvol}, one can compute the general form of the commutation relations among 
coordinates defining the type of "Lie algebra noncommutativity". One easily finds
\begin{equation}
    \left[x^M,x^\mu\right] = -i \left[\partial_{p^M}A(p^M)\rvert_{p^M=0}\right]^\mu\;_\sigma x^\sigma\;,
\label{general-liecommut}
\end{equation}
where $x^M$ is the special coordinate which generates the noncommutativity, with all other commutators being equal to $0$, which obviously satisfies the Jacobi identity so that \eqref{general-liecommut} actually defines a Lie algebra. In this paper, we are actually more interested in reversing the problem - we want to find the star products\footnote{i.e., a suitable coordinate basis, the special coordinate $x^M$ and the matrix $a(p^M)$} for the spaces from \cite{mercati}. The procedure for doing so is to find a coordinate basis\footnote{In \cite{mercati} all of the spaces are written in the Cartesian basis $(t, x, y, z)$, but the commutators in the Cartesian basis most often do not belong to the form of \eqref{general-liecommut} right away.} in which the Lie algebra is of the form \eqref{general-liecommut} and then to find the Lie group corresponding to the Lie algebra.

It turns out that \eqref{general-liecommut} encompasses all the Lie algebras listed in Section \ref{section3} as it can be easily verified. Note that in the sense of noncommutative Minkowski deformations, there exist actually three different isomorphism classes, all sharing an isomorphic Lie algebra structure \eqref{general-liecommut} whose generators are coordinate functions:
\begin{enumerate}
    \item Timelike noncommutative Minkowski spacetime deformations whose underyling Lie algebra is of the form \eqref{general-liecommut} with $x^M$ being a timelike coordinate. The matrix $A(p^M)$ then depends on the momentum in the direction of the timelike coordinate $x^M$.
    \item Spacelike noncommutative Minkowski spacetime deformations whose underlying Lie algebra is of the form \eqref{general-liecommut} with $x^M$ being a spacelike coordinate. The matrix $A(p^M)$ then depends on the momentum in the direction of the spacelike coordinate $x^M$.
    \item Lightlike noncommutative Minkowski spacetime deformations whose underlying Lie algebra is of the form \eqref{general-liecommut} with $x^M$ being a lightlike coordinate. The matrix $A(p^M)$ then depends on the momentum in the direction of the lightlike coordinate $x^M$. In the case of lightlike spaces, the usual $(t,x,y,z)$ Cartesian basis will not suffice for our considerations as none of its coordinates are lightlike. We will need to transform the basis to include the lightlike coordinate $x^M$.
\end{enumerate}

Some examples of all three classes will be given in Section \ref{section3} as the table of all applicable deformed Minkowski spaces from \cite{mercati} encompassing the respective commutator, $x^M$ and other coordinates in terms of Cartesian coordinates, the Lie algebra classification and the Lie group of the algebra and finally, the matrix $A(p^M)$ which dictates the star product behind the entire construction. It is very important to note that the three classes of noncommutative deformations of Minkowski spacetime are really distinct because it is not possible to boost or rotate a coordinate to turn it from a spatial/timelike/lightlike coordinate into one of a different type, so there isn't a justifiable way to say that symmetry of Minkowski spacetime connects any of the three distinct classes. Also, it is important to note that the relation \eqref{general-liecommut} puts a lot of constraint on basis $x^\mu$ appearing in the star product \eqref{final star-product}. Very often, the $x^M$ coordinate will not be orthogonal to other coordinates.\\

In any case, regardless of the coordinate Lie algebra's classification of the noncommutative deformation and regardless if the Lie algebra of coordinates corresponds to a space from \cite{mercati}, the star product \eqref{final star-product} gives a trace condition
\begin{equation}
    \int d^4x f\star g = \int d^4x \left[\det A(P^M)\vartriangleright g\right]\star f
\end{equation}
\begin{equation}
    P_M = -i \frac{\partial}{\partial x^M}
\end{equation}
so we can see that when $\det A \neq 1$, i.e., when the Lie group $G$ from \eqref{genesemidirect} is not unimodular, the integral is no longer a trace for the convolution algebra of functions. \\

Furthermore, for some timelike noncommutative spaces from \cite{mercati} whose corresponding Lie group is nonunimodular, an interesting modular group structure is exhibited representing a time evolution. For those groups, the Haar measure gives a "twisted trace," which is interpreted as Kubo-Martin-Schwinger (KMS) weight.  It will be discussed in the next section. 

\section{Deformed Minkowski space-times}\label{section3}
Let us first summarize in a few words the main result of Section \ref{section2}. \\
We have considered a family of quantum Minkowski space-times whose Lie groups $\mathcal{G}$ related to the Lie algebras of coordinates $\mathfrak{g}$ are semi-direct products of the form $\mathcal{G}
= H\ltimes_{{\phi}} \mathbb{R}^{n}$ where the abelian subgroup $H\subset GL(n,\mathbb{R})$ acts on $\mathbb{R}^n$ as $\phi_a(x)=ax $ for $a\in H$, $x\in\mathbb{R}^n$, $n\geq1$. We have shown that any of these quantum Minkowski space-times can be modeled by an associative $*$-algebra $\mathcal{M}=(\mathbb{C}(\mathcal{G}),\star,\dag)$ with the following star-product and involution:
\begin{equation}
    \begin{split}
        \left(f\star g\right)(x) &= \frac{1}{(2\pi)} \int dp^M dy^M e^{-i p^M y^M} f(x + y^M)g(A(p^M)x),\\
    f^\dag(x)&= \frac{1}{(2\pi)}\int dp^Mdy^M\ e^{-ip^My^M}f(A(p^M)x + y^M)
    |\det(A(p^M))|^2\label{gene-invol-bis},
    \end{split}
\end{equation}
for any $f,g\in\mathcal{M}$ and $a\in H$.\\
In this section, we will give the explicit structure of the above Lie algebras $\mathfrak{g}$ for all of the applicable Minkowski spacetime deformations from \cite{mercati}. These merely arise from the classification of the Lie algebras of dimension $2$, $3$ and $4$, see e.g. \cite{lieclassif} from which 
we borrow notation. We will consider separately the unimodular and non-unimodular cases for the group $\mathcal{G}$. In \cite{mercati} out of the 17 spaces presented, three space's Lie algebras are trivial (dubbed as space 19, 20 and 21)  when we set the central extension to zero, and three spaces (dubbed as space 9s, for $s=-1,0,1$) do not have a Lie group of the form \eqref{genesemidirect}, but rather of $\left(\mathbb{R}\ltimes\mathbb{R}\right)\ltimes \mathbb{R}^2$. All other noncommutative Minkowski deformations in \cite{mercati} are applicable to our formalism and we will use them as explicit examples later in the following subsections.\\

As some parts of this section are rather technical, we collect all the results of the Section \ref{section3} into a table, given in particular the explicit expression for the matrix $A(p^M)$ appearing in \eqref{gene-invol-bis} above. The reader not interested by the technical discussion given in the Subsections \ref{section31} and \ref{section32} may go directly to the Section \ref{section4}.\\

It is important to mention that, as explained before, we need to work in specially adapted bases which have the commutators in the form of \eqref{general-liecommut}, since the star product \eqref{final star-product} is only capable of producing commutation relations of the form \eqref{general-liecommut}. To endow the Lie algebras of noncommutative Minkowski spaces from \cite{mercati} with star products we will transform each (applicable) space's coordinate basis into a coordinate basis in which the space's commutation relations obey \eqref{general-liecommut}. To express the link between the Cartesian coordinate basis in \cite{mercati}, which we shall denote\footnote{Notice that in \cite{mercati} the coordinates are denoted as $x^0,x^1,x^2$ and $x^3$. To not cause confusion with our notation we rename $x^0=t,x^1=x, x^2=y$ and $x^3=z$ in \cite{mercati} and leave the $x^\mu$ notation for our coordinate basis which need not be orthonormal.} as $(t,x,y,z)$, and the coordinates that obey \eqref{general-liecommut}, which we shall denote as $x^\mu$, we will explicitly write every coordinate basis $x^\mu$ in terms of $(t,x,y,z)$.

\begin{table}
        \makebox[\textwidth][l]{ 
        \vspace{-1cm}
        \hspace{-2.5cm}
        \pgfplotstabletypeset[normal]{ %
         Case & Coordinates & $x^M$ &Commutators &  Group  & Matrix $A(p^M)$   \\
        & & & \textcolor{blue}{\textsc{Unimodular groups}} & & \\[-1ex]
        10 &\pbox{5cm}{$x^0 = z-t,\; x^1 = x$,\\$x^2 = z-t-y,$\\$x^3 = z.$} &$x^M = x^1$& \pbox{5cm}{$[x^M,x^2]=-i\lambda x^0$ }& $H $ & \pbox{4cm}{\tiny$\left(\begin{matrix}1 & 0 & 0 & 0\\0 & 1 & 0 & 0\\\lambda p^{1} & 0 & 1 & 0\\0 & 0 & 0 & 1\end{matrix}\right) $\normalsize} \\[3ex]
        \hline&&&&& \\[-4ex]
        11 & \pbox{5cm}{$x^0 = z-t$,\\$x^1 = x$,\;$x^2 = y$,\\ $x^3 = -z.$}&$x^M = x^2$&\pbox{5cm}{$[x^M,x^1]=-i\lambda x^0$\\
        $[x^M,x^3]= -i\lambda x^1$} & $\mathcal{G}_{4,1}$ & \pbox{4cm}{\tiny$\left(\begin{matrix}1 & 0 & 0 & 0\\\lambda p^{2} & 1 & 0 & 0\\0 & 0 & 1 & 0\\\frac{\lambda^{2} \left(p^{2}\right)^{2}}{2} & \lambda p^{2} & 0 & 1\end{matrix}\right)\normalsize
$} \\[4ex]
        \hline&&&&& \\[-4ex]
        12 & \pbox{5cm}{$x^0 = t$,\\$x^1 = x$,\;$x^2 = y$,\\ $x^3 = t-z.$} &$x^M = x^1$&\pbox{5cm}{$[x^M,x^0]=-i\lambda x^3 $ }& $H $ & \pbox{4cm}{\tiny$\left(\begin{matrix}1 & 0 & 0 & \lambda p^1\\0 & 1 & 0 & 0\\0 & 0 & 1 & 0\\0 & 0 & 0 & 1\end{matrix}\right) $\normalsize} \\[3ex]
        \hline&&&&& \\[-4ex]
        13& \pbox{5cm}{$x^0 = t,\; x^1 = x$,\\$x^2 = y,\;x^3 = z.$} &$x^M = x^0$&\pbox{5cm}{$[x^M,x^1]=i \lambda x^2$ \\$ [x^M,x^2]=-i \lambda x^1$} &$SE(2) $ & \pbox{4cm}{\tiny$\left(\begin{matrix}1 & 0 & 0 & 0\\0 & \cos{\left(\lambda p^{0} \right)} & - \sin{\left(\lambda p^{0} \right)} & 0\\0 & \sin{\left(\lambda p^{0} \right)} & \cos{\left(\lambda p^{0} \right)} & 0\\0 & 0 & 0 & 1\end{matrix}\right)$\normalsize} \\[3ex]
        \hline&&&&& \\[-4ex]
        14 & \pbox{5cm}{$x^0 = t,\; x^1 = x$,\\$x^2 = y,\;x^3 = z.$} &$x^M = x^3$&\pbox{5cm}{$[x^M,x^1]=i \lambda x^2$ \\$ [x^M,x^2]=-i \lambda x^1$} &$SE(2) $ &\pbox{4cm}{\tiny$\left(\begin{matrix}1 & 0 & 0 & 0\\0 & \cos{\left(\lambda p^{3} \right)} & - \sin{\left(\lambda p^{3} \right)} & 0\\0 & \sin{\left(\lambda p^{3} \right)} & \cos{\left(\lambda p^{3} \right)} & 0\\0 & 0 & 0 & 1\end{matrix}\right)$\normalsize}\\[3ex]
        \hline&&&&& \\[-4ex]
        15 & \pbox{5cm}{$x^0 = t+z$,\\$x^1 = x$,\;$x^2 = y$,\\ $x^3 = z.$} &$x^M = x^0$&\pbox{5cm}{$[x^M,x^1]=i \lambda x^2$ \\$ [x^M,x^2]=-i \lambda x^1$} &$SE(2) $ & \pbox{4cm}{\tiny$\left(\begin{matrix}1 & 0 & 0 & 0\\0 & \cos{\left(\lambda p^{0} \right)} & - \sin{\left(\lambda p^{0} \right)} & 0\\0 & \sin{\left(\lambda p^{0} \right)} & \cos{\left(\lambda p^{0} \right)} & 0\\0 & 0 & 0 & 1\end{matrix}\right)$\normalsize}\\[3ex]
        \hline&&&&& \\[-4ex]
        16 & \pbox{5cm}{$x^0 = t,\; x^1 = x$,\\$x^2 = y,\;x^3 = z.$} &$x^M = x^1$&\pbox{5cm}{$[x^M,x^0]=i\lambda x^3 $\\$ [x^M,x^3]=i\lambda x^0$ }& $SE(1,1) $ & \pbox{5cm}{\tiny$\left(\begin{matrix}\cosh{\left(\lambda p^{1} \right)} & 0&0&-\sinh{\left(\lambda p^{1} \right)} \\0 & 1 &0 & 0 \\0 & 0 & 1 & 0\\ -\sinh{\left(\lambda p^{1} \right)} & 0 & 0 & \cosh{\left(\lambda p^{1} \right)}\end{matrix}\right)$\normalsize} \\[4ex]
        \hline\hline & & & \textcolor{blue}{\textsc{Nonunimodular groups}} & & \\
        \pbox{5cm}{7 \\$(\alpha = \frac{1}{\zeta})$} &\pbox{5cm}{$x^0 = \alpha t$,\\$x^1 = x,\; x^2 = y$,\\ $x^3 = \alpha(z - t)$} &$x^M = x^0$& \pbox{5cm}{$[x^M,x^1]=i\lambda(\alpha x^1 + x^2),$\\$[x^M,x^2]=i\lambda(\alpha x^2-x^1)$\\ $[x^M,x^3]=i\lambda\alpha x^3$} & $\mathcal{G}_{4,6}^{\alpha, \alpha}$ & \hspace{-10mm} \pbox{4cm}{\tiny$\left(\begin{matrix}1 & 0& 0 & 0\\0 & e^{- \alpha \lambda p^{0}} & 0 & 0\\0 & 0 & e^{- \alpha \lambda p^{0}} \cos(\lambda p^{0}) & -e^{- \alpha \lambda p^{0}} \sin(\lambda p^{0})\\0 & 0 & e^{- \alpha \lambda p^{0}} \sin(\lambda p^{0}) & e^{- \alpha \lambda p^{0}} \cos(\lambda p^{0})\end{matrix}\right)$\normalsize}\\[4ex]
        \hline&&&&& \\[-3ex]
        8& \pbox{5cm}{$x^0 = t$,\\$x^1 = x,\; x^2 = y$,\\ $x^3 = \zeta(t - z)$} &$x^M = x^0$&\pbox{8cm}{$[x^M,x^1]=i\lambda(x^3 + x^1)$\\$ [x^M,x^2]=i\lambda x^2$\\$ [x^M,x^3]=i\lambda x^3$}& $\mathcal{G}_{4,2}^{1}$ &  \pbox{4cm}{\tiny$\left(\begin{matrix}1 & 0 & 0 & 0\\0 & e^{- \lambda p^{0}} & 0 & - \lambda p^{0} e^{- \lambda p^{0}}\\0 & 0 & e^{- \lambda p^{0}} & 0\\0 & 0 & 0 & e^{- \lambda p^{0}}\end{matrix}\right)$\normalsize}\\[4ex]
        \hline&&&&& \\[-4ex]
        17 & \pbox{5cm}{$x^0 = -t$,\\$x^1 = x,\;x^2 = y$,\\ $x^3 = z-t$} &$x^M = x^0$&$[x^M,x^3] = i\lambda x^3$& $\mathcal{G}_{2,1}$ & \pbox{4cm}{\tiny$\left(\begin{matrix}1 & 0 & 0 & 0\\0 & 1 & 0 & 0\\0 & 0 & 1 & 0\\0 & 0 & 0 & e^{- \lambda p^{0}}\end{matrix}\right)$\normalsize}\\[3ex]
        \hline&&&&& \\[-4ex]
        \pbox{5cm}{18 \\$\alpha = \frac{1}{\zeta}$} &\pbox{5cm}{$x^0 =\alpha t$,\\$x^1 = x-y$,\\$x^2 = x+y$,\\ $x^3 =t- z.$} & $x^M = x^0$&\pbox{8cm}{$[x^M,x^1]=i\lambda x^2$\\$ [x^M,x^2]=-i\lambda x^1$\\$ [x^M,x^3]=i\lambda x^3$}& $\mathcal{G}_{4,6}^{1, 0}$ &   \pbox{4cm}{\tiny$\left(\begin{matrix}1 & 0 & 0 & 0\\0 & \cos{\left(\lambda p^{0} \right)} & - \sin{\left(\lambda p^{0} \right)} & 0\\0 & \sin{\left(\lambda p^{0} \right)} & \cos{\left(\lambda p^{0} \right)} & 0\\0 & 0 & 0 & e^{- \lambda p^{0}}\end{matrix}\right)$\normalsize}\\
}}
\caption{\textbf{Classification of Lie algebras.} 11 Minkowski spacetime deformations are described. The numbers in the left column come from Zakrzewski numeration \cite{zakrzew}, the next column gives the explicit coordinate change to map Mercati \cite{mercati} bracket to the \eqref{general-liecommut} form of the Lie algebra. Next, the commutator relations and the associated Lie groups are given. The last column gives the $A(p^M)$ matrix, with parameter $p^M$ encoding the time/light/space-like noncommutativity.}
\end{table}

\subsection{Coordinates Lie algebras with unimodular groups}\label{section31}
In this subsection we will deal with 3- and 4-dimensional Lie algebras. Note that the 3-dimensional algebras can actually be adapted to a 4-d quantum space-time by merely enlarging the initial set of coordinates with an additional central coordinate. 
\subsubsection{3-d Lie algebras}\label{section311}

Typical 3-d Lie algebras illustrating the present framework correspond to the cases 13, 14, 15 of \cite{mercati}, \cite{zakrzew}. The corresponding non trivial part of the Lie algebras{\footnote{These algebras involve a central coordinate which can be incorporated at the end of the construction.}} can be easily verified to be isomorphic to the Lie algebra $\mathfrak{iso}(2)$. \\
The related Lie group is the Euclidean group $E(2)\simeq O(2)\ltimes \mathbb{R}^2$ which, upon focusing only on orientation preserving isometries, reduces to the special Euclidean group 
\begin{equation}
    SE(2)\simeq SO(2)\ltimes \mathbb{R}^2\label{euclideangroup}
\end{equation}
which is linked to the $\rho$-Minkowski space-time as shown in \cite{rho-weyl}. This deformation of the Minkowski space-time has been introduced a long ago in \cite{lukier1} and investigated more recently in \cite{lizzi1}, \cite{lizzi2}.\\
The star-product and associated involution have been derived recently in \cite{rho-weyl} using the general construction of Section \ref{section2}. One merely combines \eqref{gene-invol-bis} to the faithful representation of \eqref{euclideangroup}, 
$\gamma:(a,\vec{x})
    \longmapsto \begin{pmatrix} a & \vec{x} \\ 0 & 1   \end{pmatrix} $, 
    $\vec{x}\in\mathbb{R}^2$, where the matrix $a$ can be easily cast into
\begin{equation}
a(p^M)=\begin{pmatrix}\cos(\rho p^M)&\sin(\rho p^M)\\-\sin(\rho p^M)&\cos(\rho p^M)  \end{pmatrix}\label{rotation-matrix}
\end{equation}
where we have restored the deformation parameter $\rho$ which has the dimension of a length. The resulting expressions are trivially obtained from \eqref{gene-invol-bis} so that we do not reproduce them here. We note again that describing a 4-d situation can be easily obtained by incorporating a central coordinate (see e.g. \cite{rho-weyl}).\\

The case 16 of ref.\cite{mercati} where all the central extension $\theta$ parameters have been set to zero can be viewed as a Lorentzian version of the above cases.
The related relevant Lie group is $E(1,1)\simeq SO(1,1)\ltimes \mathbb{R}^2$ with faithful representation \eqref{decadix-gamma} characterized by the matrix
\begin{equation}
    a(p^3)=\begin{pmatrix}\cosh(\lambda p^3)&\sinh(\lambda p^3)\\\sinh(\lambda p^3)&\cosh(\lambda p^3),\label{decadix-11}
    \end{pmatrix}
\end{equation}
where we have again restored a parameter $\lambda$ with length dimension.\\
The relevant star-product and involution are then readily obtained from \eqref{decadix-11} combined with \eqref{gene-invol-bis}. For this 3-d version of the deformed Minkowski space-time, one easily obtains
\begin{equation}
  (f\star g)(x^3,\vec{x})=\int dy^3dp^3\ e^{-iy^3p^3}f(x^3+y^3,\vec{x})g(x^3,u^0,u^1),\label{star-iso11}
\end{equation}
\begin{equation}
    f^\dag(x^3,\vec{x})=\int dy^3dp^3\ e^{-iy^3p^3}f(x^3+y^3,u^0,u^1)
\end{equation}
with
\begin{equation}
u^0=x^0\cosh(\lambda p^3)+x^1\sinh(\lambda p^3),\ u^1=x^0\sinh(\lambda p^3)+x^1\cosh(\lambda p^3)
\end{equation}
which yields the following coordinates algebra   
\begin{equation}
   [x^3,x^0]=-i\lambda x^1,\  [x^3,x^1]=-i\lambda x^0, \ [x^0,x^1]=0,
\end{equation}
while the replacement of \eqref{decadix-11} by \eqref{rotation-matrix} into \eqref{gene-invol-bis} would produce the (3-d) coordinates algebra for $\rho$-Minkowski given by 
\begin{equation}
   [x^0,x^1]=+i\lambda x^2,\  [x^0,x^2]=-i\lambda x^1, \ [x^1,x^2]=0,
\end{equation}
as explained above.\\

Finally, a similar analysis holds for the case 10 and 12 with the parameter $\theta$ set to $0$ for which the algebra is found to be isomorphic to the Heisenberg algebra. The corresponding matrix $a(\lambda p^1)$ characterizing the associated Lie group is now given by
\begin{equation}
a(\lambda p^1)=\begin{pmatrix}1&\lambda p^1\\
0&1
\end{pmatrix}
.
\end{equation}

\subsubsection{4-d Lie algebras}\label{section312}
A typical 4-d Lie algebra of coordinates is provided with the case 11 in \cite{mercati} with $\theta=0$. The corresponding algebra is
\begin{align}
[t,x]&=0,\ [t,y]=-ix,\ [t,z]=0,\nonumber\\
[x,y]&=i(z-t),\ [y,t]=ix,\ [x,z]=0\label{ggg}.
\end{align}
making the substitution $x^0 = z-t$ , one verifies that \eqref{ggg}  is isomorphic to the indecomposable Lie algebra $\mathfrak{g}_{4,1}$ of \cite{lieclassif}, namely
\begin{equation}
    [x^2,x^0]=-i\lambda x^1,\ [x^2,x^1]=-i\lambda x^0\label{g41},
\end{equation}
while all the other commutators vanish. Note that this algebra is one of the only two nilpotent 4-d Lie algebras.\\
The related Lie group can be again characterized as in eqn. \eqref{decadix-gamma} with the matrix $a(p^2)$ now given by \cite{lieclassif}
\begin{equation}
    a(\lambda p^2)=\begin{pmatrix}
        1&\lambda p^2&\lambda^2\frac{(p^2)^2}{2}\\
        0&1&\lambda p^2\\
        0&0&1
    \end{pmatrix}\label{g41-matrix}.
\end{equation}
The expression for the star-product and involution can then be obtained as a simple application of \eqref{gene-invol-bis} combined with \eqref{g41-matrix}.\\

\subsection{Coordinates Lie algebras with non-unimodular groups}\label{section32}

\subsubsection{2-d Lie algebras}
The space defined by case (17) is a 2-d Lie algebra which corresponds to 2-d light like $\kappa$ Minkowski with 2 commutative coordinates.
\subsection{4-d Lie algebras}
The first Lie algebra of coordinates we consider in this subsection is the one associated to a non-unimodular group and corresponds to the case 7 of \cite{zakrzew} (see also \cite{mercati}). It can be written as
\begin{align}
    [t,x]&=i(x+\zeta u),\ [t,y]=i(y-\zeta x),\ [t,z]=i(z-t)\nonumber,\\
    [y,z]&=-i(y-\zeta x),\ [t,x]=i(x+\zeta y),\ [x,y]=0,\label{cas7}
\end{align}
where $\zeta>0$ is a dimensionless parameter. Note that the case $\zeta=0$ corresponds to the coordinate algebra for the usual $\kappa$-Minkowski space-time as it can be easily verified.\\
This Lie algebra is isomorphic to the Lie algebra $\mathfrak{g}_{4,6}^{\alpha,\beta}$ \cite{lieclassif} with $\alpha=\frac{1}{\zeta}$, as it can be easily seen by using the following change of generators into \eqref{cas7}
\begin{equation}
x^0=\alpha t,\ x^1=x,\ x^2=y,\ x^3=\alpha(z-t)\label{change-cas7}
\end{equation}
leading to
\begin{equation}
[x^0,x^1]=i (\alpha x^{1} + x^{2}),\ [x^0,x^2]= i(\alpha x^2 - x^1),\ [x^0,x^3]=i \alpha x^{3},\label{g46}
\end{equation}
while the other commutators vanish, which define $\mathfrak{g}_{4,6}^{\alpha,\alpha}$ \cite{lieclassif} upon redefining the generators as 
\begin{equation}
E_k=ix^k,\ k=1,2,3,\ E_4=ix^0. \label{phys-redef}
\end{equation}
The final coordinates Lie algebra takes the form
\begin{equation}
[x^0,x^1]=i \lambda(\alpha x^{1} + x^{2}),\ [x^0,x^2]=\lambda(\alpha x^2 - x^1),\ [x^0,x^3]=i \lambda \alpha x^{3},
\label{g46-final}
\end{equation}
with $\lambda$ (as always) having the dimension of length, which models a specific "noncommutativity" among the coordinates of a quantum Minkowski space-time.\\

This 4-dimensional Lie algebra is indecomposable. It turns out that the Lie groups corresponding to 4-dimensional indecomposable algebras have been classified\footnote{Along with all other Lie algebras up to, and including, the dimension $4$.} in \cite{lieclassif}. From this work, one infers that the Lie group related to \eqref{matrixg46-final} can be represented \cite{lieclassif} as in \eqref{decadix-gamma}, namely any element of the group has the matrix form
\begin{equation}
\begin{pmatrix} a(\lambda p^0) & \vec{x} \\ 0 & 1   \end{pmatrix},\ \vec{x}=(x_1,x_2,x_3),\label{matrixg46}
\end{equation}
with
\begin{equation}
a(\lambda p^0)=\begin{pmatrix} e^{-\alpha\lambda p^0} & 0 & 0 \\ 0 &  e^{-\alpha\lambda p^0}\cos(\lambda p^0)& -e^{-\alpha\lambda p^0}\sin(\lambda p^0)\\
0&e^{-\alpha\lambda p^0}\sin(\lambda p^0)&e^{-\alpha\lambda p^0}\cos(\lambda p^0) \end{pmatrix}\label{matrixg46-final}.
\end{equation}
One checks that $\det(a(p_0))=e^{-3\alpha\lambda p^0}$ signaling that the group $\mathcal{G}$ is not-unimodular in view of \eqref{detexplcit}. Then, from \eqref{matrixg46-final} and \eqref{gene-invol-bis}, one obtains the expression of the star-product together with the involution for the quantum Minkowski space-time with coordinates algebra \eqref{g46-final}. They are given by
\begin{equation}
    (f\star g)(x^0,\vec{x})=\int dy^0dp^0\ e^{-iy^0p^0}f(x^0+y^0,\vec{x})g(x^0,u^1,u^2,e^{-\alpha\lambda p^0}x^3)\label{star-g46},
\end{equation}

\begin{equation}
    f^\dag(x^0,\vec{x})=\int dy^0dp^0\ e^{-iy^0p^0} e^{-6\alpha\lambda p^0}f(x^0+y^0,u^1,u^2,e^{-\alpha\lambda p^0}x^3)\label{invol-g46},
\end{equation}
with
\begin{equation}
u^1=e^{-\alpha\lambda p^0}(x^1\cos(\lambda p^0)-x^2\sin(\lambda p^0)),\
u^2=e^{-\alpha\lambda p^0}(x^2\cos(\lambda p^0)+x^1\sin(\lambda p^0)).
\end{equation}
One easily verifies that the commutation relations \eqref{g46-final} can be recovered from \eqref{star-g46}.\\

One remark is in order here. The case 18 in \cite{mercati} can be treated in a similar way. It can be verified that the Lie algebra of coordinates is isomorphic to the algebra $\mathfrak{g}_{4,6}^{\alpha,\beta}$ \cite{lieclassif} with $\alpha=\frac{1}{\zeta}$, $\beta=0$. In this latter case, the matrix $a$ takes the form
\begin{equation}
    a(\lambda p^0)=\begin{pmatrix} e^{-\alpha\lambda p^0} & 0 & 0 \\ 0 &  \cos(\lambda p^0)& -\sin(\lambda p^0)\\
0&\sin(\lambda p^0)&\cos(\lambda p^0) \end{pmatrix}.
\end{equation}
The corresponding star-product and involution can be obtained from e.g. \eqref{star-g46}, \eqref{invol-g46} by obvious modifications.\\

Another Lie algebra associated to a non-unimodular group is related to the case 8 of \cite{zakrzew}, \cite{mercati}. The corresponding commutation relations are
\begin{align}
[t,x]&=i(x-\zeta(z-t)),\ [x,z]=-i(x-\zeta(z-t)),\nonumber\\
[t,z]&=i(z-t),\ [t,y]=iy,\nonumber\\
[y,z]&=-iy,\ [x,y]=0\label{cas8}
\end{align}
where again $\zeta>0$ is a dimensionless parameter. Upon defining 
\begin{equation}
    x^0=-t\ x^1=x,\ x^3=\zeta(z-t),\ x^2=y\label{change-cas8}
\end{equation}
one realizes that the Lie algebra \eqref{cas8} is isomorphic to the indecomposable Lie algebra $\mathfrak{g}_{4,2}^1$, see e.g. \cite{lieclassif} whose defining commutation relations are
\begin{equation}
    [x^2,x^0]=-ix^2,\ [x^1,x^0]=-i(x^3+x^1),\ [x^3,x^0]=-ix^3\label{g42},
\end{equation}
while the other commutators vanish. Then, combining \eqref{g42} 
with \eqref{phys-redef}, one obtains
\begin{equation}
[x^0,x^1]=i\lambda x^1,\ [x^0,x^2]=i\lambda x^2,\ [x^0,x^3]=i\lambda (x^2+x^3)\label{g42-final}
\end{equation}
which describes the noncommutativity among the coordinates of another quantum Minkowski space-time which we will now characterize.\\

The Lie group $\mathcal{G}$ associated with \eqref{g42-final} can be represented \cite{lieclassif} in the matrix form similar to the one given in \eqref{matrixg46} with however $a(\lambda p_0)$ given by
\begin{equation}
    a(\lambda p^0)=\begin{pmatrix} e^{-\lambda p^0} & 0 & 0 \\ 0 &  e^{-\lambda p^0}& 0\\
-\lambda p^0e^{-\lambda p^0}&0&e^{-\lambda p^0} \end{pmatrix}.
\end{equation}
This group is not unimodular since $\det(a(p^0))=e^{-3\lambda p^0}$. The star-product and the related involution can be easily obtained from the analysis of Section \ref{section2}. We find
\begin{equation}
    (f\star g)(x^0,\vec{x})=\int dy^0dp^0\ e^{-iy^0p^0}f(x^0+y^0,\vec{x})g(x^0,e^{-\lambda p^0}x^1,u^2,e^{-\lambda p^0}x^3)\label{star-g46},
\end{equation}

\begin{equation}
    f^\dag(x^0,\vec{x})=\int dy^0dp^0\ e^{-iy^0p^0} e^{-6\lambda p^0}f(x^0+y^0,e^{-\lambda p^0}x^1,u^2,e^{-\lambda p^0}x^1)\label{invol-g46},
\end{equation}
with
\begin{equation}
   u^2=e^{-\lambda p_0}(x^2-\lambda p^0x^3).
\end{equation}

\subsection{KMS weight as a twisted trace for non-unimodular groups}

For non-unimodular groups it appears that the Lebesgue measure actually defines a "twisted trace". A twisted trace (on an algebra) can be roughly defined as a linear positive map, denoted as $\tr$, satisfying $\tr(ab) = \tr((\sigma\triangleright b)a)$, where $\sigma$ is an automorphism of the algebra, called the twist. Actually, the name "twisted trace" is a slight abuse of language and should be {\it{stricto sensu}} replaced by "KMS weight" whose characterization will be given below.  Because all of the spaces with nonunimodular group structure that we explicitly considered in this paper are in the time noncommutative class, the results of this subsection will be applicable to all of them. In this subsection we will denote $x^M = x^0$ and $x^\mu = \vec{x}$.\\

A simple computation yields
\begin{equation}
    \int dx^0 d\vec{x}\; (f\star g)(x)=\int dx^0 d\vec{x} dy^0dp^0\ {\det A(p^0)}\ e^{-iy^0p^0}g(x^0 + y^0,\vec{x})f(A(p^0)x)\label{masterform}
\end{equation}
for any $a(p^M)\in H$ and $f,g\in \mathcal{M}$ where we used \eqref{matrix-exponent}, \eqref{flip-inverse} and 
\eqref{star-prodetinvol}.\\
One then infers that whenever the group $\mathcal{G}$ is unimodular, which is verified provided $\det(A(p^0))=1$,
\eqref{masterform} reduces obviously to
\begin{equation}
     \int dx^0 d\vec{x}\ (f\star g)(x)= \int dx^0 d\vec{x}\ (g\star f)(x)\label{cyclicity}
\end{equation}
for any $f,g\in \mathcal{M}$ so that one concludes that the usual Lebesgue integral is cyclic w.r.t the star product \eqref{star-prodetinvol}. Furthermore, one can verify that for any $f,g\in \mathcal{M}$ 
\begin{equation}
    \int dx^0 d\vec{x}\ (f^\dag\star g)(x)=\int dx^0 d\vec{x}\ \overline{f}(x)g(x)\label{positivemap}
\end{equation}
provided $\det(A(p^0))=1$ still holds. From \eqref{positivemap} upon setting $f=g$, one concludes that the Lebesgue integral $\int dx^0 d\vec{x}$ defines a positive map $\int dx_0 d\vec{x}:\mathcal{M}_+\to\mathbb{R}^+$ where $\mathcal{M}_+$ denotes the positive elements of $\mathcal{M}$ which moreover is cyclic w.r.t the star-product from \eqref{cyclicity}. Hence, $\int dx^0 d\vec{x}$ defines a trace when the group $\mathcal{G}$ is unimodular as announced above.\\

A convenient way to begin with the case when $\mathcal{G}$ is not unimodular is to use the explicit expressions of $\det(A(p^0))$ for each of the cases considered in this paper. As it can be verified from Section \ref{section3}, one obtains generically (setting again $\lambda = 1$ for the time being)
\begin{equation}
    \det(A(p_0))=e^{-np_0}\;,
    \label{detexplcit}
\end{equation}
where $n$ depends on the dimension of the Lie algebra of coordinates (see Section \ref{section3}), namely, $n=1$ or $n=3$. Then, the combination of \eqref{detexplcit} with \eqref{masterform} gives
\begin{equation}
    \int dx^0d\vec{x}\ (f\star g)(x^0,\vec{x})= \int dx^0d\vec{x}\ ((\sigma\triangleright g)\star f)(x^0,\vec{x})\label{twistedtrace-gene},
\end{equation}
where the twist $\sigma\in\textrm{Aut}(\mathcal{M})$ is given by
\begin{equation}
    (\sigma\triangleright g)(x^0,\vec{x})=(e^{in\partial_0}\triangleright g)(x^0,\vec{x})=g(x^0+in,\vec{x})\label{twist-gene}
\end{equation}
for any $f,g\in\mathcal{M}$. Observe by the way that the Lebesgue integral is nothing but the right Haar measure of $\mathcal{G}$, as it can be easily verified 
from \eqref{modul-function1}, \eqref{measure1} combined with the general relation $d\mu_\mathcal{G}(s)=\Delta_\mathcal{G}(s) d\nu_\mathcal{G}(s)$ for any $s\in\mathcal{G}$ where $d\nu_\mathcal{G}$ denotes the right Haar measure. One can easily check that
\begin{equation}
    (\sigma\triangleright f)^\dag=\sigma^{-1}\triangleright (f^\dag)\label{regular}
\end{equation}
thus defining a regular automorphism \cite{connes-mosco}. Note that a quite similar twist equipping a "twisted trace" has already been shown to occur within 
the $\kappa$-Minkowski space-time \cite{kappa-weyl-bis} as the corresponding group of the coordinates algebra is known to be the non-unimodular semi-direct product $\mathbb{R}\ltimes \mathbb{R}^n$ whose structure fits 
with \eqref{genesemidirect}. \\

Related to \eqref{twist-gene}, a distinguished one-parameter group of $*$-automorphims of $\mathcal{M}$ is defined by
\begin{equation}
 \{\sigma_t:=e^{tn\partial_0} \}_{t\in\mathbb{R}},  \label{tomita-group}
\end{equation}
with obviously $\sigma_{t=i}=\sigma$ and satisfies
\begin{equation}
    \sigma_{t_1}\sigma_{t_2}=\sigma_{t_1+t_2},\ \sigma_{t}^{-1}=\sigma_{-t},
\end{equation}
\begin{equation}
    \sigma_{t}\triangleright(f\star g)=(\sigma_{t}\triangleright f)\star 
    (\sigma_{t}\triangleright g),\ (\sigma_{t}\triangleright f)^\dag=(\sigma_{t}\triangleright f^\dag),
\end{equation}
for any $f,g\in\mathcal{M}$ and $t,t_1,t_2\in\mathbb{R}$.\\
Moreover, a simple computation gives
\begin{equation}
    \int dx^0d\vec{x}\ (f\star g^\dag)(x^0,\vec{x})= \int dx^0d\vec{x}\ f(x^0,\vec{x})\overline{g}(x^0,\vec{x})\label{quasi-closedness},
\end{equation}
for any $f,g\in\mathcal{M}$, which is true for both the unimodular and nonunimodular groups $\mathcal{G}$ in our formalism. This latter relation insures that the Lebesgue integral always defines a positive map $\int dx^0d\vec{x}:\mathcal{M}_+\to\mathbb{R}^+$.\\

It turns out that the group of $*$-automorphisms \eqref{tomita-group} is the modular group for the KMS weight defined by the above map, with Tomita operator given by $\Delta_T=e^{-in\partial_0}$, since one has $\sigma_t=(\Delta_T)^{it}$. The action of any $\sigma_t$ \eqref{twist-gene} on $\mathcal{M}$ generates time translations in view of
\begin{equation}
    (\sigma_t\triangleright f)(x_0,\vec{x})=f(x_0+int,\vec{x}),
\end{equation}
which can be read off from \eqref{twist-gene}. This actually represents a time evolution for the operators related to the Weyl quantization map \eqref{Weyl-map}. Note that this supports {\it{a posteriori}} our choice in the labeling of coordinates.\\

The characterization of the Lebesgue integral as a KMS weight is exactly the same as the one presented in \cite{kappa-weyl-bis}. Recall that the (positive) map $\Phi$, defined by the Lebesgue integral, is a KMS weight if it satisfies the two conditions 
\begin{equation}
    \Phi(\sigma_z\triangleright f)=\Phi(f),\ \Phi(f\star f^\dag)=\Phi\big((\sigma_{\frac{i}{2}}\triangleright f) \star (\sigma_{\frac{i}{2}}\triangleright f)^\dag\big )\label{kms-definition}
\end{equation}
for any $f\in\mathcal{M}$ where $\sigma_z$, $z\in\mathbb{C}$, analytically extends $\sigma_t$ \eqref{tomita-group} with now $\sigma_z(f^\dag)=(\sigma_{\overline{z}}(f))^\dag$. It appears that these conditions hold true stemming from standard computation. Then a theorem (see e.g. theorem 6.36 of ref. \cite{kustermans}) guarantees the existence of a bounded continuous function $F:\Sigma\to\mathbb{C}$ with $\Sigma=\{z\in\mathbb{C}, 0\le\Im(z)\le1 \}$ such that $F(t)=\Phi((\sigma_t\triangleright f)\star g)$, $F(t+i)=\Phi(g\star (\sigma_t\triangleright f))$ which can be formally identified to the KMS condition.\\

\section{Related deformations of Poincar\'e symmetry}\label{section4}

Before concluding, let us briefly touch on the question of the symmetry Hopf algebras of the explored NC Minkowski spaces. Take the following action for the generators of the Poincar\'e alegbra
\begin{equation}
P_\mu\triangleright f=-i\partial_\mu f,\ \ M_{\mu\nu}\triangleright f=i(x_\mu\partial_\nu-x_\nu\partial_\mu)f\;,\label{cestlavraieaction}
\end{equation}
of course, for such an action \eqref{cestlavraieaction}, one obtains the usual Lie algebra structure
\begin{align}
    [P_\mu,P_\nu]&=0,\ [M_{\mu\nu},P_\rho]=\eta_{\nu\rho}P_\mu-\eta_{\mu\rho}P_\nu,\nonumber\\
    [M_{\mu\nu},M_{\rho\tau}]&=\eta_{\nu\rho}M_{\mu\tau}-\eta_{\mu\rho}M_{\nu\tau}-\eta_{\nu\tau}M_{\mu\rho}+\eta_{\mu\tau}M_{\nu\rho}\;.
\end{align}
We can search for the associated coproduct $\Delta$ which is compatible with the aformentioned action \eqref{cestlavraieaction} and our star product:
\begin{equation}
    \mu \circ \Delta(h) (f\otimes g) = h\triangleright (f\star g),\quad \forall h\in \mathcal{P}_\lambda\;.
\label{coproduct definition}
\end{equation}
The definition \eqref{coproduct definition} is part of a general procedure of building a deformed symmetry Hopf algebra from the duality with the deformed $\mathcal{C}^*$ algebra . This procedure, however, when considering the star products for the spaces from \cite{mercati}, which should have Poincar\'e Hopf algebras of deformed symmetry\footnote{It is well known that in the Drinfeld twist approach one is able to find a twisted Poincar\'e Hopf algebra for all of the considered cases from \cite{mercati}, so we expect the same to happen for our formalism.}, produces $i\mathfrak{gl}(1,3)$ Hopf algebras\footnote{In other words, for those cases the coproduct \eqref{coproduct definition} does not exist if only allowing the generators of the Poincar\'e algebra.} for all the cases except for 13,14,15 and 16, which are shown below in the respective bases from the Table 1 (Here $C = \cos(\lambda P_M)$ and $S = \sin(\lambda P_M)$): 
\begin{equation*}
\hspace{-17mm}
\setlength{\arraycolsep}{0.5em}
\begin{array}{l|l}
\begin{split}
\text{Space} & \;13\\
        \Delta{\left(M_{10} \right)}=&M_{10} \otimes 1+\lambda P_{1}  \otimes M_{21}+C\otimes M_{10}+S\otimes M_{20}\\
\Delta{\left(M_{20} \right)}=&M_{20} \otimes 1+\lambda P_{2}  \otimes M_{21}+C\otimes M_{20}+S \otimes M_{01}\\
\Delta{\left(M_{21} \right)}=&1 \otimes M_{21}+M_{21} \otimes 1\\
\Delta{\left(M_{30} \right)}=&1 \otimes M_{30}+M_{30} \otimes 1+\lambda P_{3}  \otimes M_{21}\\
\Delta{\left(M_{31} \right)}=&M_{31} \otimes 1+C \otimes M_{31}+S \otimes M_{32}\\
\Delta{\left(M_{32} \right)}=&M_{32} \otimes 1+C\otimes M_{32}+S \otimes M_{13}
\end{split}
&
\begin{split}
\text{Spaces }&14,15\\
       \Delta{\left(M_{10} \right)}=&M_{10} \otimes 1+C\otimes M_{10}+S \otimes M_{20}\\
\Delta{\left(M_{20} \right)}=&M_{20} \otimes 1+C\otimes M_{20}+S \otimes M_{01}\\
\Delta{\left(M_{21} \right)}=&1 \otimes M_{21}+M_{21} \otimes 1\\
\Delta{\left(M_{30} \right)}=&1 \otimes M_{30}+M_{30} \otimes 1+\lambda P_{0} \otimes M_{21}\\
\Delta{\left(M_{31} \right)}=&M_{31} \otimes 1+\lambda P_{1} \otimes M_{21}+C \otimes M_{31}+S\otimes M_{32}\\
\Delta{\left(M_{32} \right)}=&M_{32} \otimes 1+\lambda P_{2} \otimes M_{21}+C \otimes M_{32}+S \otimes M_{13}
\end{split}
\end{array}
\end{equation*}

\begin{center}
    
\begin{equation*}
\hspace{-17mm}
\setlength{\arraycolsep}{0.5em}
\begin{array}{l}
\begin{split}
\text{Space} & \;16\\
\Delta{\left(M_{10} \right)}=&\;M_{10} \otimes 1+\lambda P_{0} \otimes M_{30}+\cosh{\left(\lambda P_{M} \right)} \otimes M_{10}+\sinh{\left(\lambda P_{M} \right)} \otimes M_{31}\\\Delta{\left(M_{20} \right)}=&\;M_{20} \otimes 1+\cosh{\left(\lambda P_{1} \right)} \otimes M_{20}+\sinh{\left(\lambda P_{1} \right)} \otimes M_{32}\\\Delta{\left(M_{21} \right)}=&\;1 \otimes M_{21}+M_{21} \otimes 1+\lambda P_{2} \otimes M_{03}\\\Delta{\left(M_{30} \right)}=&\;1 \otimes M_{30}+M_{30} \otimes 1\\\Delta{\left(M_{31} \right)}=&\;M_{31} \otimes 1+\lambda P_{3} \otimes M_{03}+\cosh{\left(\lambda P_{M} \right)} \otimes M_{31}+\sinh{\left(\lambda P_{M} \right)} \otimes M_{10}\\\Delta{\left(M_{32} \right)}=&\;M_{32} \otimes 1+\cosh{\left(\lambda P_{M} \right)} \otimes M_{32}+\sinh{\left(\lambda P_{M} \right)} \otimes M_{20}
\end{split}
\end{array}
\end{equation*}
\end{center}
For the spaces 13, 14 and 15, in the respective change of basis defined in the Table 1, one obtains using the described method the above coproduct structures and they are equivalent to (4.28) in \cite{lizzi2} up to possible renaming of coordinates. Additionally, it turns out that the same construction is applicable also to the case 16, the "hyperbolic" $\rho$-Minkowski spacetime, with the coproduct structure as displayed above  and one can see that this new result is very similar to the already known case of the $\rho$-Minkowski. 

An important remark is that in this approach, unlike in the Drinfeld twist approach, the relation \eqref{coproduct definition} is not guaranteed by any underlying principles. It is not, so far, ruled out that our convolutional star product produces a noncommutative Minkowski spacetime for which the quantum symmetry Hopf algebra, no matter the action $\triangleright$, can not close in the Poincar\'e algebra and satisfy \eqref{coproduct definition} at the same time. Perhaps, alongside with a modification of the $\triangleright$ from the trivial one to a nontrivial one, one also needs to generalize the relation \eqref{coproduct definition} itself to include some $\lambda$ depending terms.
On the other hand, it is also not ruled out that the relation \eqref{coproduct definition} can be satisfied in all of the spaces and the failure to close in the Poincar\'e algebra can be completely attributed to the inappropriate action of the generators \eqref{cestlavraieaction}. There may exist, for all cases, a nonlinear change in the expression of generators which could relate the convolutional and Drinfeld twist formalisms. For example, this was achieved for the timelike and spacelike $\kappa$ Minkowski. With similar convolutional star products, the action used for the convolutionally defined $\kappa$ Minkowski spacetimes in \cite{kappa-weyl-bis} actually employed the Poincar\'e generator action found earlier in \cite{lukier2}. In any case, it seems that for the spaces 13-16, both the generators' actions and the relation \eqref{coproduct definition} are valid. Finally, in the Drinfeld twist approach, \cite{tolstoy2007} computed the twisted coproducts $\Delta_\mathcal{F}$ for the spaces with abelian $r$ matrices of \cite{Z2}, and gave the procedure to compute the coproducts for nonabelian $r$ matrices describing the spaces from \cite{mercati}. \\

\section{Discussion and conclusion} \label{section5}

Let us summarize the main points of this paper. \\

We have constructed star-products and related involutions characterizing the $*$-algebras of Minkowski spaces with Lie algebras of coordinates belonging to a broad family of Lie algebras. This formalism encompasses 11 quantum Minkowski space-times derived from \cite{zakrzew}, \cite{mercati}, using for that purpose the properties of convolution algebras for the Lie groups related to the coordinates Lie algebras of these quantum space-times combined with the Weyl quantization. The quantum space-times we have explicitly considered correspond to the non-centrally extended Lie algebras of coordinates out of the 17 classes of quantum space-times derived in \cite{mercati}. \\
These star-products and involutions can be expressed through rather simple expressions, roughly speaking Fourier transforms of convolution products, so that they appear to be conveniently designed to be used in the construction of field theories on these so far poorly explored quantum Minkowski space-times. \\

When the Lie algebra of coordinates corresponds to a unimodular Lie group, we have shown that the usual Lebesgue integral is a natural choice for a trace, while for non-unimodular groups, the notion of trace must be traded for a "twisted trace," which is known in the mathematical literature as a KMS weight. Note that the appearance of a KMS weight \cite{kustermans} has already been exploited in field theories and gauge theories on $\kappa$-Minkowski spaces \cite{MW2020a}-\cite{tadpole}. Recall that the KMS weight is an important tool in the Tomita-Takesaki modular theory \cite{takesaki} which plays a salient role in the overall set-up of the interesting thermal time hypothesis \cite{connes-rov}, \cite{connes-rov2} which is still under debate. For a very recent related study within $\kappa$-Minkowski space-time extending the analysis \cite{kappa-weyl-bis}, see \cite{kilian-therm}. Note that all the twists defining the various KMS weights arising in the present analysis are similar to the one showing up in \cite{MW2020a}, as it is apparent from \eqref{detexplcit}, to be compared to the expression for the twist of \cite{MW2020a}, \cite{kilian-therm}. It would be interesting to examine if some physical features could emerge from the various KMS weights arising in some of the quantum space-times considered in this paper. \\

There exist several ways to build a deformed symmetry Hopf algebra $\mathcal{P}_\lambda$ that reduces at the commutative limit to the Poincaré algebra. One of them is based on Drinfeld twists\footnote{Given a Hopf algebra $\mathcal{H}$, a Drinfeld twist is an invertible element $\mathcal{F} \in \mathcal{H} \otimes \mathcal{H}$ satisfying normalisation and $2$-cocycle conditions.} and it is suitable for noncommutative spaces built with such twists. Such spaces have star products given by 
\begin{equation}
f \star_{\mathcal{F}} g = \mu \circ \mathcal{F} \triangleright(f \otimes g)
\end{equation}
and their deformed symmetry Hopf algebras have deformed coproducts 
\begin{equation}
    \Delta^{\mathcal{F}} = \mathcal{F} \circ \Delta \circ \mathcal{F}^{-1}\;.
\label{Drinfeld coproduct}
\end{equation}
The antipode $S^\mathcal{F}$ is also given with the canonical, undeformed, antipode $S$ and with the Drinfeld element $\mathcal{F}$, while the rest of the Hopf algebra $\mathcal{P}_\lambda $ structure is undeformed\footnote{Note that this approach can, in some cases, lead to $  i\mathfrak {gl}(1,3)$ Hopf algebra's deformation. This is the case when the generator in the twist involves dilatonic (or other $i\mathfrak{gl}$) generators, as can happen for $\kappa$-Minkowski. In this paper, however, we only explicitly consider deformations of spaces with $r$-matrices from \cite{zakrzew} which only contain generators of the Poincaré algebra. Thus, for the 11 Lie algebras from \cite{mercati} we expect Poincaré deformed Hopf algebras, not $i\mathfrak{gl}(1,3)$.}. The deformed coproduct obtained in this way is, trivially from \eqref{Drinfeld coproduct}, compatible with the star product $\star_{\mathcal{F}}$ in the sense of \eqref{coproduct definition}:
\begin{equation}
    h \triangleright(f \star_{\mathcal{F}} g) = \mu_{\star_{\mathcal{F}}} \circ \Delta_{\mathcal{F}}(h)(f \otimes g), \quad \forall h \in \mathcal{P}_\lambda\;.
    \label{compatibility}
\end{equation}
As mentioned in Section IV, the procedure from \cite{tolstoy2007} can be used to construct twisted Hopf algebras for all spaces from \cite{mercati}. However, in our formalism, as our star product has not been built using a Drinfeld twist, compatibility relation \eqref{compatibility} is not guarenteed to hold anymore and the correct Poincar\'e action on a case by case basis is unknown for spaces \cite{mercati}. Using the undeformed action of Poincaré generators \eqref{cestlavraieaction} on $\mathcal{M}_\lambda$ and supposing \eqref{coproduct definition} in many cases yields a coproduct which is not closed with respect to the Poincare algebra\footnote{For \cite{mercati}'s spaces, in all spaces, except for 13,14,15 and 16, this failiure to close in Poincar\'e algebra happens.}. We keep the results for spaces $13-16$ for which the closure happens. Other spaces need to be investigated in a future work. \\


The present framework involves a relevant material to construct and study easily classical and quantum properties of field theories and gauge theories built on various interesting quantum Minkowski space-times which are almost unexplored so far. This opens the way for related systematic studies which we will present in forthcoming publications.\\

\bibliographystyle{unsrt}

\end{document}